\documentclass[12pt]{article}
\usepackage[margin=0.8in]{geometry}
\usepackage[utf8]{inputenc}
\usepackage{graphicx}
\usepackage{amssymb}
\usepackage{amsmath}
\usepackage{amsfonts}
\usepackage{xcolor}
\usepackage{url}
\usepackage{float}
\usepackage{soul}
\usepackage{hyperref}
\hypersetup{
  colorlinks   = true, 
  urlcolor     = blue, 
  linkcolor    = black, 
  citecolor   = black 
}

\usepackage[numbers,sort&compress]{natbib}
\let\OLDthebibliography\thebibliography
\renewcommand\thebibliography[1]{
  \OLDthebibliography{#1}
  \setlength{\parskip}{4pt}
  \setlength{\itemsep}{0pt plus 0.3ex}
}

\newcommand{\tautau}{{\tau^+\tau^-}}
\newcommand{\gamgam}{{\gamma\gamma}}
\newcommand{\bbbar}{{b \bar b}}
\newcommand{\gev}{\,\mathrm{GeV}}
\newcommand{\mev}{\,\mathrm{MeV}}
\newcommand{\tev}{\,\mathrm{TeV}}
\newcommand{\tb}{{\tan\beta}}
\newcommand\mh[1]{m_{h_{#1}}}
\newcommand\SW{s_\mathrm{w}}
\newcommand\CW{c_\mathrm{w}}
\newcommand\MW{M_W}
\newcommand\MZ{M_Z}
\newcommand\MWSM{\MW^{\rm SM}}
\newcommand\MWNTHDM{\MW^{\rm N2HDM}}
\newcommand\MWCDFnew{\MW^{\rm CDF-new}}
\newcommand\sweff{\sin^2 \theta_\mathrm{eff}}

\newcommand\al{\alpha}
\newcommand\be{\beta}

\newcommand\De{\Delta}
\newcommand\si{\sigma}
\newcommand\MHp{m_{H^\pm}}

\newcommand{\mt}{m_t}
\newcommand{\he}{\ensuremath{h_{95}}}
\newcommand{\hz}{\ensuremath{h_{125}}}

\newcommand\refeq[1]{Eq.~(\ref{#1})}

\newcommand\refse[1]{Sect.~\ref{#1}}

\newcommand\citere[1]{Ref.~\cite{#1}}
\newcommand\citeres[1]{Refs.~\cite{#1}}
\def\reffi#1{\mbox{Fig.~\ref{#1}}}

\newcommand\LB{\left[}
\newcommand\RB{\right]}
\newcommand\LV{\left\{}
\newcommand\RV{\right\}}

\begin{document}

\thispagestyle{empty}

\def\thefootnote{\fnsymbol{footnote}}

\begin{flushright}
  DESY 22-067 ~~\\ IFT-UAM/CSIC--22--043 ~~
\end{flushright}

\vspace*{1.4cm}

\begin{center}

{\Large \textbf{Excesses in the low-mass Higgs-boson search 
    and the\\[.5em] \boldmath{$W$}-boson mass measurement}}

\vspace{1cm}

{\sc
T.~Biekötter\footnote{thomas.biekoetter@desy.de}$^1$,
S.~Heinemeyer\footnote{Sven.Heinemeyer@cern.ch}$^{2}$
and  G.~Weiglein\footnote{georg.weiglein@desy.de}$^{1,3}$
}

\vspace*{.7cm}

{\sl
$^1$Deutsches Elektronen-Synchrotron DESY,
Notkestr.~85,  22607  Hamburg,  Germany

\vspace{0.1cm}

$^2$Instituto de F\'isica Te\'orica UAM-CSIC, Cantoblanco, 28049,
      Madrid, Spain

\vspace{0.1cm}

$^3$II.\  Institut f\"ur  Theoretische  Physik, Universit\"at  Hamburg,
Luruper Chaussee 149, 22761 Hamburg, Germany
}

\end{center}

\vspace*{0.1cm}

\begin{abstract}
The CDF collaboration recently reported a measurement 
of the $W$-boson
mass $M_W$
whose value shows
a large upward deviation from 
the Standard Model
(SM) prediction.
The question arises whether 
such large values of~$M_W$
could be accommodated in extensions of 
the SM without violatong other constraints and which  
phenomenological
consequences this would have.
A different type of deviation from the SM has been observed
experimentally in the searches for light
Higgs bosons. CMS has observed
two 
excesses with a local significance
of about $3\,\sigma$
in the $\gamma\gamma$ and $\tau^+\tau^-$ final states for a
hypothetical Higgs-boson mass of 
approximately $95$~GeV.
These two excesses are
compatible with the corresponding ATLAS limits.
Earlier an excess was
also observed in the Higgs-boson searches at LEP in the 
$b \bar b$ final state at the local
$2\,\sigma$ confidence level at about
the same 
mass.
It was shown recently that the three
excesses can be accommodated in 
a Two-Higgs-Doublet Model that is extended with a real singlet (N2HDM) 
of Yukawa type~IV, while being
in agreement with all other theoretical and experimental constraints.
We demonstrate here that the
region of the parameter space that describes the three
excesses can also give rise to a large contribution to $M_W$ in agreement with
the recent CDF measurement. We discuss further phenomenological
consequences of this scenario.
\end{abstract}

\def\thefootnote{\arabic{footnote}}
\setcounter{page}{0}
\setcounter{footnote}{0}

\newpage

\section{Introduction}
\label{sec:int}
The mass of the $W$~boson
$M_W$ can be predicted from muon decay, which relates
$\MW$ to three extremely precisely measured quantities: the Fermi
constant $G_\mu$, the fine structure constant $\al$,
and the mass of the
$Z$~boson $\MZ$. Within the Standard Model (SM) and many extensions of it,
this relation can be used to predict $\MW$ via the expression
\begin{align}
\MW^2 = \MZ^2 \LV \frac{1}{2} +
\sqrt{\frac{1}{4} - \frac{\pi\,\al}{\sqrt{2}\,G_\mu\,\MZ^2}
\LB 1 + \De r(\MW, \MZ, \mt, \ldots) \RB } \RV~,
\label{eq:mwpred}
\end{align}
where the quantity $\De r$ is zero at lowest order.\footnote{
For an example of a model where the
lowest-order prediction for $\MW$ is modified, see 
for instance \citere{Diessner:2019ebm}.}
It comprises loop corrections to muon decay in the considered model, where the
ellipsis in \refeq{eq:mwpred} denotes the specific particle content of the
model.

The known contributions to the SM prediction for $\De r$ include the complete
one-loop~\cite{Sirlin:1980nh,Marciano:1980pb} and the complete
two-loop result~\cite{Djouadi:1987gn,Djouadi:1987di,Kniehl:1989yc,Halzen:1990je,Kniehl:1991gu,Kniehl:1992dx,Halzen:1991ik,Freitas:2000gg,Freitas:2002ja,Awramik:2002wn,Awramik:2003ee,Onishchenko:2002ve,Awramik:2002vu,Bauberger:1996ix,Bauberger:1997ey,Awramik:2006uz},
as well as partial higher-order corrections up to four-loop
order~\cite{Avdeev:1994db,Chetyrkin:1995ix,Chetyrkin:1995js,Chetyrkin:1996cf,Faisst:2003px,vanderBij:2000cg,Boughezal:2004ef,Schroder:2005db,Chetyrkin:2006bj,Boughezal:2006xk}.
This yields a prediction of
\begin{align}
\MWSM = 80.353 \gev~,
\label{mwsm}
\end{align}
where the implementation of \citere{Stal:2015zca} in the code {\tt FeynHiggs}~\cite{Heinemeyer:1998yj,Hahn:2009zz,Bahl:2018qog} has been used, 
and the input parameters have been chosen as in \citere{Bagnaschi:2022qhb}.
The remaining uncertainty from unknown higher-order corrections has been estimated 
in \citere{Awramik:2003rn} to be about $4 \mev$
(see also \citere{Degrassi:2014sxa}).
Concerning the theoretical uncertainty that is
induced by the experimental errors of the input parameters,
in particular $M_Z$ and the top-quark mass $m_t$ are relevant.
A variation of $M_Z$ within its $\pm 1\,\sigma$ interval
yields a shift of $\pm 2.7 \mev$, while a variation of
$m_t$ by $\pm 0.5 \gev$ yields a shift of about $\pm 3 \mev$.
The SM prediction of \refeq{mwsm} is somewhat lower
than the current
PDG average of the experimental results
prior to the new
CDF measurement~\cite{ParticleDataGroup:2020ssz},
\begin{align}
\MW^{\rm PDG} &= 80.379 \pm 0.012 \gev\, ,
\label{exp-cwa}
\end{align}
but in agreement at the level
of about $2 \, \sigma$.

Recently the CDF collaboration reported a new measurement using their
full data set of $8.8$\,fb$^{-1}$~\cite{CDF:2022hxs}, 
\begin{align}
\MWCDFnew &= 80.4335 \pm 0.0094 \gev\,,
\label{cdf-new}
\end{align}
which deviates from the SM prediction by about $7\,\si$. 
The CDF collaboration also reported a combination of this new measurement 
with the other results from the Tevatron and with the measurements at LEP
(but not with the results from the LHC),
\begin{align}
\MW^{\rm Tev+LEP} &= 80.4242 \pm 0.0087 \gev\,.
\label{tev-lep}
\end{align}
In the future it will be mandatory to assess the compatibility
of the different measurements of $\MW$ with each other and to carefully analyze 
possible sources of systematic effects.
The inclusion of the CDF measurement into a future world average is expected to give rise to an upward shift of the central value. 
Accordingly, the
question arises whether extensions of the SM exist that can
accommodate values between the central value of the current world average 
and the value measured by CDF without being in conflict with existing
experimental and theoretical constraints, and which further phenomenological
consequences arise from such scenarios. The implications of this new measurement
for various scenarios of physics beyond the SM (BSM) have been discussed
in \citere{deBlas:2022hdk,Yang:2022gvz,Strumia:2022qkt,
Yuan:2022cpw,Athron:2022qpo,Lu:2022bgw,Fan:2022dck,
Babu:2022pdn,Heckman:2022the,Gu:2022htv,
Athron:2022isz,DiLuzio:2022xns,Asadi:2022xiy,
Bahl:2022xzi,Paul:2022dds,Bagnaschi:2022whn,
Song:2022xts,Cheng:2022jyi,Lee:2022nqz,
Liu:2022jdq,Fan:2022yly,Zhu:2022tpr,Sakurai:2022hwh}.

\smallskip
Theories that could potentially
accommodate a large contribution to
$M_W$ that shifts the predicted value towards
$M_W^{\rm CDF-new}$
might also feature other direct signals of new
physics that are detectable at the LHC.
While so far no conclusive signs of 
BSM physics
have been found at the LHC,
both the measurements of the properties
of the Higgs
boson at $125\gev$
(its couplings are known up to now
to an experimental precision of roughly
10 to $20\%$~\cite{CMS:2022dwd,ATLAS:2022vkf}) 
and the existing limits from the searches for new particles
leave ample room for BSM physics
at (or below) the EW scale.
In this paper we will study the
possibility of an extended Higgs sector
in which (as will be shown below)
a value of $M_W$ compatible with the
CDF measurement can be realized by means
of sizable corrections to $\Delta r$
arising from mass splittings between the BSM Higgs bosons that have sizable
couplings to the gauge bosons.
A direct way of experimentally probing
such a possibility consists of direct
searches for the additional 
Higgs bosons at colliders, which
have been performed at
LEP~\cite{Abbiendi:2002qp,Barate:2003sz,Schael:2006cr},
the Tevatron~\cite{Group:2012zca} and
the LHC~\cite{CMS:2017yta,Sirunyan:2018aui,
Sirunyan:2018zut,ATLAS:2018xad,CMS-PAS-HIG-21-001}.
In this context one should note that, even though
no detection of additional Higgs bosons has been made
so far, several intriguing excesses in the searches
for light Higgs bosons below $125~\gev$ have been observed.

Results based on the first year of
CMS~Run~2 data for Higgs-boson
searches in the diphoton
final state show a local excess of about~$3\,\sigma$ at a mass of
$95\gev$~\cite{Sirunyan:2018aui}, where
a similar excess of~$2\,\sigma$ occurred in
the Run~1 data at a comparable
mass~\cite{CMS:2015ocq}.
Combining 7, 8~and first year $13 \tev$ data
(and assuming that the $gg$ production dominates)
the excess is most pronounced at a mass
of $95.3 \gev$ with a local significance of $2.8\,\sigma$.
First Run\,2~results from~ATLAS
with~$80$\,fb$^{-1}$ in the~$\gamgam$~searches below~$125$\,GeV were
reported in 2018~\cite{ATLAS:2018xad}.
Although
no significant excess above the~SM~expectation was
observed in the mass range between
$65$~and $110 \gev$,
the limit on cross section times branching ratio obtained in the
diphoton final state by ATLAS is substantially weaker than the corresponding upper limit obtained by CMS
at and around $95 \gev$.

CMS recently published the results for
the search for additional Higgs bosons in the $\tautau$
channel~\cite{CMS-PAS-HIG-21-001}. 
Utilizing the full Run~2 data set
the CMS collaboration
reported an excess in the low-mass region
assuming the gluon-fusion production
mode and the subsequent decay into $\tautau$ pairs,
where the mass range of the excess
is compatible with the 
excesses that have been observed
by CMS in the diphoton searches
during Run~1 and Run~2.
The excess in the $\tautau$ final state is
most pronounced for a mass hypothesis of
$100\gev$, with a local significance of
$3.1\,\sigma$,
while for a mass value of $95\gev$,
i.e.~closer to the most significant 
excess in the $\gamgam$ search~\cite{Sirunyan:2018aui}, 
CMS reports a local significance of $2.6\,\sigma$.
So far there exists no corresponding 
result for the low-mass search in the
$\tautau$ final state from the ATLAS 
collaboration in this mass range.

Searches for a low-mass Higgs boson that were previously carried out at 
LEP resulted in a $2.3\,\sigma$ local excess
observed in the~$e^+e^-\to Z(H\to b\bar{b})$
process~\cite{Barate:2003sz}
at a mass of
about $98 \gev$, where due to the
$b \bar b$ final state the
mass resolution was rather coarse.
Because of this limited mass resolution in the
$\bbbar$ final state at LEP  
this excess can be compatible with the slightly lower mass of $95 \gev$
favoured by the
CMS excesses.

Since the reported excesses in the $\gamgam$ and $\tautau$ channels
at the LHC  and the $\bbbar$ channel at LEP
were found at approximately the same mass, 
the question of a possible common origin arises. Recently we
demonstrated that indeed all three excesses can be described
consistently in the N2HDM (the Two-Higgs-Doublet Model with an
additional real singlet~\cite{Chen:2013jvg,Muhlleitner:2016mzt}) of
Yukawa type~IV~\cite{Biekotter:2022jyr}.
\footnote{Analyses in the N2HDM
(and extensions)
focusing only
on the $\gamgam$ and $\bbbar$
excesses can be found in \citeres{Biekotter:2019kde,Biekotter:2019mib,Biekotter:2019drs,Biekotter:2020ahz,Biekotter:2020cjs,Biekotter:2021ovi,Heinemeyer:2021msz,Biekotter:2021qbc}.}
~In the present analysis we demonstrate that the parameter space that
accommodates the three 
excesses can also give rise to a large contribution 
to $\MW$ that can even bring the 
predicted value into agreement with the central value of
the recent CDF measurement
(as discussed above, a future world average for
$M_W$ including the CDF measurement would be
expected to lie in between the current world
average and the central value reported by~CDF).
We discuss further phenomenological
consequences of this scenario.

The paper is organized as follows.
After introducing the model in \refse{sec:n2hdm}
and the relevant
theoretical and experimental constraints on
the N2HDM parameter space in \refse{sec:constraints},
the numerical results of our parameter scan are presented in
\refse{sec:numana}, where also the future prospects are discussed.
We summarize our results in \refse{sec:conclu}.

\section{Model definition, relevant constraints and the prediction for 
\boldmath{$\MW$}}
\label{sec:model}

In the following we give a brief description of the N2HDM
in order to define the necessary quantities and to introduce our notation.
More details can be
found in \citere{Biekotter:2022jyr}.

\subsection{The N2HDM}
\label{sec:n2hdm}

The N2HDM is the simplest
extension of a CP-conserving Two-Higgs doublet model (2HDM)
in which the latter is augmented with a real
scalar singlet Higgs 
field~\cite{Chen:2013jvg,Muhlleitner:2016mzt}.
As in the 2HDM we define $\tb := v_2/v_1$,
the ratio of the 
vacuum expectation values (vev)
of the two $SU(2)$ doublet fields
$\Phi_1$ and $\Phi_2$.
In order
to avoid the occurrence of tree-level flavor-changing neutral currents
(FCNC), a $Z_2$ symmetry is imposed
under which either $\Phi_1$ or $\Phi_2$
changes the sign, and
which is only softly
broken by a bilinear term 
$m_{12}^2 ( \Phi_1^\dagger \Phi_2 + \mathrm{h.c.})$.
As in the 2HDM, one can have four variants of
the N2HDM, depending on the $Z_2$ parities of the
fermions. We will focus on type~IV
(flipped), which was shown to 
be capable of accommodating the three excesses at $95 \gev$.
In addition, the scalar potential is invariant under a second
$Z_2$ symmetry acting only on $\Phi_S$.
This symmetry is spontaneously broken if the singlet
acquires a vev $v_S$.
The CP-even scalar spectrum contains a total of three
physical Higgs bosons
$h_{1,2,3}$, where we use the convention
$\mh1 < \mh2 < \mh3$.
The relation between the 
states in the gauge eigenstate basis
and the physical states
can be expressed in
terms of the
$3 \times 3$ orthogonal matrix~$R$,
which can be parameterized by
three mixing angles 
$- \pi / 2 \leq \alpha_1, \alpha_2, \alpha_3
\leq \pi / 2$
such that
\begin{equation}
R=
\begin{pmatrix}
c_{\alpha_1}c_{\alpha_2} &
  s_{\alpha_1}c_{\alpha_2} &
    s_{\alpha_2} \\
-(c_{\alpha_1}s_{\alpha_2}s_{\alpha_3}+s_{\alpha_1}c_{\alpha_3}) &
  c_{\alpha_1}c_{\alpha_3}-s_{\alpha_1}s_{\alpha_2}s_{\alpha_3}  &
    c_{\alpha_2}s_{\alpha_3} \\
-c_{\alpha_1}s_{\alpha_2}c_{\alpha_3}+s_{\alpha_1}s_{\alpha_3} &
-(c_{\alpha_1}s_{\alpha_3}+s_{\alpha_1}s_{\alpha_2}c_{\alpha_3}) &
c_{\alpha_2}c_{\alpha_3}
\end{pmatrix} \ ,
\label{mixingmatrix}
\end{equation}
where we use the short-hand
notation $s_x = \sin x$, $c_x = \cos x$.

The couplings of the Higgs bosons to
the fermions and gauge bosons
are modified
w.r.t.~to the couplings of
a Higgs boson as predicted by the SM.
We express the couplings of the scalar
mass eigenstates $h_i$, normalized to the corresponding SM couplings,
in terms of the coupling coefficients $c_{h_i V V}$ and
$c_{h_i f \bar f}$, such that the couplings
to the massive vector bosons
are given by
\begin{equation}
\left(g_{h_i W W}\right)_{\mu\nu} =
\mathrm{i} g_{\mu\nu} \left(c_{h_i V V}\right) g M_W
\quad \text{and } \quad
\left(g_{h_i Z Z}\right)_{\mu\nu} =
\mathrm{i} g_{\mu\nu} \left(c_{h_i V V}\right) \frac{g M_Z}{\CW} \, ,
\end{equation}
where $g$ is the $SU(2)_L$ gauge coupling,
$\CW = \MW/\MZ$ is the cosine of the weak
mixing angle, and $\SW = \sqrt{1 - \CW^2}$.
The couplings of the Higgs bosons
to the SM fermions are given by
\begin{equation}
g_{h_i f \bar{f}} =
\frac{m_f}{v} \left(c_{h_i f \bar{f}}\right) \; ,
\end{equation}
where $m_f$ is the mass of the fermion, and
$v = \sqrt{(v_1^2 +v_2^2)} \approx 246\gev$ is the SM vev.
Analytical expressions for these coupling coefficients in terms of the
mixing angles $\al_{1,2,3}$ and $\be$ can be found
in~\citere{Biekotter:2019kde}.

The scalar potential of the N2HDM comprises 11
free parameters.
We use the public code \texttt{ScannerS}~\cite{Coimbra:2013qq,
Muhlleitner:2016mzt,Muhlleitner:2020wwk},
with which the model
can be explored in terms of the parameters
\begin{equation}
c^{2}_{h_{2}t\bar{t}} \, , \;\;
c^{2}_{h_{2}VV} \, , \;\;
\mathrm{sign}(R_{23}) \, , \;\;
R_{13} \, , \;\;
\tb \, , \;\;
v_{S} \, , \;\;
m_{h_{1,2,3}} \, , \;\;
m_{A} \, , \;\;
\MHp \, , \;\;
m_{12}^{2} \, . \label{eq:inputsnew}
\end{equation}
Here, $m_A$, $\MHp$ denote
the masses of the physical CP-odd and charged Higgs bosons, respectively.
We will identify the lightest CP-even Higgs boson
$h_1$ with the one
that could
potentially be identified with the excesses at
$95 \gev$, labelled \he.
The second-lightest CP-even Higgs boson
$h_2$ will be
identified with the detected
state at $125 \gev$, labelled \hz.
Besides the 11 free parameters mentioned above,
\refeq{eq:inputsnew} also contains the 
input parameter
$\mathrm{sign}(R_{23})$, which is used
to lift a degeneracy
arising from the dependence of
the mixing angles $\alpha_i$
on the squared values of the coupling
coefficients $c^{2}_{h_{2}t\bar{t}}$
and $c^{2}_{h_{2} VV}$ and the
element of the mixing matrix $R_{13}$.

\subsection{Theoretical and experimental constraints}
\label{sec:constraints}

In our analysis we apply several theoretical requirements 
to the parameter space of the N2HDM. We give here only a very
brief description; more details can be found in
\citere{Biekotter:2022jyr}.

In order to ensure that for the considered parameter point
the electroweak minimum
is physically viable it
is required that the EW vacuum
is either stable or meta-stable,
i.e.~sufficiently long-lived in comparison
to the age of the universe.
In particular, we apply conditions on the scalar couplings
that exclude parameter points for which the
scalar potential is not bounded from
below~\cite{Klimenko:1984qx,Muhlleitner:2016mzt}.
For the calculation of the
lifetime of the EW vacuum
in case the electroweak minimum
is not the global minimum of the potential,
\texttt{ScannerS} provides an
interface to the public code
\texttt{EVADE}~\cite{Hollik:2018wrr,Ferreira:2019iqb}.
We also apply the
tree-level perturbative unitarity conditions that
ensure that in the high-energy limit
the eigenvalues of the
scalar $2\times 2$ scattering matrix are
smaller than $|8\pi|$~\cite{Muhlleitner:2016mzt}.

The experimental constraints are applied as follows.
We verify the agreement of the selected points with
the currently available measurements of the properties of the Higgs boson at
about 125~GeV using the public code
\texttt{HiggsSignals
v.2.6.1}~\cite{Bechtle:2013xfa,Stal:2013hwa,
Bechtle:2014ewa,Bechtle:2020uwn}.
The required theoretical
predictions for the cross sections and the branching
ratios of the scalars are obtained from the public
codes \texttt{SusHi}~\cite{Harlander:2012pb,Harlander:2016hcx}
and \texttt{N2HDECAY}~\cite{Djouadi:1997yw,
Butterworth:2010ym,Djouadi:2018xqq,
Muhlleitner:2016mzt,Engeln:2018mbg}.
In the following we denote as $\chi^2_{125}$
the $\chi^2$ contribution obtained from \texttt{HiggsSignals}.
We demand that each
point fulfills the condition
\begin{equation}
\chi^2_{125} - \chi^2_{125,\mathrm{SM}} \leq 5.99 \,. 
\label{eq:chi125constraint}
\end{equation}
This corresponds
to an exclusion limit on the model parameters
in a joint estimation of two parameters
at the 95\% confidence level under the
assumption that the SM fit result
$\chi^2_{125,\mathrm{SM}}$ 
provides a good approximation for
the best-fit $\chi^2_{125}$-value of the
N2HDM (see \citere{Bechtle:2020uwn} for
details).\footnote{We checked that for
all points $\chi^2_{125} > \chi^2_{125,\mathrm{SM}}$.
Thus, the application of the condition shown
in \refeq{eq:chi125constraint} is more restrictive than
demanding that
no parameter point is disfavoured at more
than $95\%$ confidence level compared to
the best-fit point regarding $\chi^2_{125}$.}
We have checked explicitly that
this is the case for
our parameter scan.
In order to test the parameter points
against the exclusion limits from the Higgs-boson
searches at LEP, the
Tevatron and in particular from the LHC,
we employ the public code
\texttt{HiggsBounds v.5.9.1}~\cite{Bechtle:2008jh,Bechtle:2011sb,
Bechtle:2013gu,Bechtle:2013wla,Bechtle:2015pma,Bechtle:2020pkv}.
Constraints from flavor-physics observables
are taken into account by
the approach as implemented in \texttt{ScannerS},
where the 2HDM flavor constraints
projected to the $\tan\beta$--$m_{H^\pm}$ plane
as given in \citere{Haller:2018nnx}
are applied 
as approximation for the N2HDM, see e.g.\ the discussion in 
\citere{Biekotter:2021qbc}.
In particular, large tensions arise
in 2HDM-like extensions of the SM for small
values of $\tan\beta$ for the measurements of
leptonic and radiative $B$-meson
decays~\cite{Haller:2018nnx}.
We will therefore use $\tan\beta = 1$ as
a lower limit in our parameter scans.
Contrary to previous analyses, we do not
apply constraints from
electroweak precision observables (EWPOs).
Instead, we investigate whether agreement with the
Tevatron measurement of~$M_W$ can be achieved, and we discuss the extent of
compatibility with the experimental results for the effective leptonic weak
mixing angle at the $Z$-boson reesonance,
see the next subsection.

\subsection{Prediction for \boldmath{$\MW$}}
\label{sec:mw}

If new physics contributions to the EWPOs
enter mainly through gauge boson self-energies, as it is
the case for extended Higgs sectors,
the BSM effects in the predictions for $\MW$ and the $Z$-pole observables
can in a simple
approximation be expressed in terms of
the oblique parameters $S$, $T$ and
$U$~\cite{Peskin:1990zt,Peskin:1991sw}.
We make use of the implementation 
of the one-loop contributions to the oblique
parameters for the N2HDM 
in \texttt{ScannerS}, which is based on
generic results for extended Higgs sectors with an
arbitrary number of Higgs doublets
and singlets from~\citere{Grimus:2008nb}.
In the framework of the
oblique parameters,
the $W$-boson mass can be calculated
using the expression~\cite{Grimus:2008nb}
\begin{equation}
\MW^2 = \left( M_W^2 \right)^{\rm SM}
\left(
1 + \frac{\SW^2}{\CW^2 - \SW^2} \Delta r'
\right) \ ,
\label{MW-STU}
\end{equation}
with
\begin{equation}
\Delta r' = \frac{\alpha}{\SW^2}
\left(
-\frac{1}{2} S +
\CW^2 T +
\frac{\CW^2 - \SW^2}{4 \SW^2} U
\right) \ .
\end{equation}
Inserting the results for $S$, $T$ and $U$ obtained 
with \texttt{ScannerS} in the N2HDM yields our
prediction for the mass of
the $W$ boson, in the following denoted by~$\MWNTHDM$.

Another very precisely known EWPO
is the leptonic effective weak mixing angle at the $Z$-boson resonance,
usually referred to as $\sweff$.
$\MW$ and $\sweff$ are the observables that by far have the largest impact
on the electroweak fit, and we therefore do not include further $Z$-pole
observables in our analysis. In fact, for the 
total width of the $Z$ boson, $\Gamma_Z$,
it was shown in \citere{Bahl:2022xzi} that the tension
between the experimental value of
$\Gamma_Z$ and its theoretical
prediction is only at the level of $1\sigma$
in the range of $T$ that is required to
predict a value of $M_W$ in agreement with
the CDF measurement.

In order to investigate the corresponding shift
in $\sweff$ that is induced via 
the BSM effects in terms of the oblique parameters
as a consequence of sizable contributions to
the prediction for $\MW$, we
compute $\sweff$ according to~\cite{Ciuchini:2013pca}
\begin{equation}
\sin^2\theta_{\rm eff} =
\sin^2\theta_{\rm eff}^{\rm SM} - 
\frac{\alpha (S - 4 \CW^2 \SW^2 T)}{4(\CW^2 - \SW^2)} \ .
\label{eq:sweffT}
\end{equation}
As will be shown below,
the numerically relevant contribution
arises from the
$T(= \Delta \rho / \alpha)$ parameter.
For the values of the SM parameters that
enter in the prediction of
$M_W^{\rm SM}$ and $\sin^2\theta_{\rm eff}^{\rm SM}$
we used the
set of numerical values as given
in Eq.~(7) of \citere{Bagnaschi:2022qhb}, which were taken from
\citeres{ParticleDataGroup:2020ssz,
Steinhauser:1998rq}. The SM 
predictions for $\MW$ and $\sweff$ obtained in this way are
$\MW^{\rm SM} = 80.353 \gev$ and $\sweff^{\rm SM} = 0.23156$.

\subsection{Fitting the excesses at 95 GeV}
\label{sec:fit95}

In order to analyze whether
the N2HDM can yield an upward shift in the prediction
for $\MW$ that is sufficiently large to make it 
compatible with the CDF measurement 
and simultaneously provide a possible 
explanation of
the observed 
$\gamgam$, $\tautau$ and $\bbbar$ excesses,
we perform a $\chi^2$-analysis
quantifying the agreement between the
theoretically predicted signal rates and
the experimentally observed values.
Experimentally, it was determined that the
excesses at $95\gev$ were best described assuming
signal rates of a scalar resonance of
\begin{align}
\mu_{\gamma\gamma}^{\rm exp} \pm \Delta
  \mu_{\gamma\gamma}^{\rm exp} &= 0.6 \pm
    0.2~\text{\cite{Sirunyan:2018aui}} \ , \\
\mu_{bb}^{\rm exp} \pm \Delta
  \mu_{bb}^{\rm exp} &= 0.117 \pm
    0.057~\text{\cite{Abbiendi:2002qp}} \ , \\
\mu_{\tau\tau}^{\rm exp} \pm \Delta
  \mu_{\tau\tau}^{\rm exp} &= 1.2 \pm
    0.5~\text{\cite{CMS-PAS-HIG-21-001}} \ ,
\end{align}
where the signal strengths are defined
as the cross sections times branching ratios
divided by the corresponding predictions
for a hypothetical SM Higgs boson at
the same mass, and the experimental uncertainties
are given as $1\ \sigma$ uncertainties.
The theoretically predicted
values $\mu_{\gamma\gamma}$, $\mu_{bb}$
and $\mu_{\tau\tau}$ were obtained by
computing the gluon-fusion production cross
section of $h_{95}$ with the help of
\texttt{SusHi}~\cite{Harlander:2012pb,Harlander:2016hcx},
and the branching ratios for the Higgs bosons
were computed using
\texttt{N2HDECAY}~\cite{Djouadi:1997yw,
Butterworth:2010ym,Djouadi:2018xqq,
Muhlleitner:2016mzt,Engeln:2018mbg}
(see \citere{Biekotter:2022jyr}
for more details).
For each individual excess, we define
the $\chi^2$ 
contributions
\begin{equation}
\chi^2_{\gamgam,\tau\tau,bb} =
\frac{
(\mu_{\gamgam,\tau\tau,bb} -
\mu_{\gamgam,\tau\tau,bb}^{\rm exp})^2 }{
(\Delta \mu_{\gamgam,\tau\tau,bb}^{\rm exp})^2} \ .
\end{equation}
In order to 
assess the combined description
of the three excesses, we define the
total $\chi^2$ contribution as
\begin{equation}
\chi^2_{\gamma\gamma+\tau\tau+bb} =
\chi^2_{\gamma\gamma} + \chi^2_{\tau\tau} +
\chi^2_{bb} \ , 
\end{equation}
where the 
results for the three 
channels
in which the
excesses were observed are treated as
independent measurements, such that we can
simply add the three individual
$\chi^2$ contributions.
In the following numerical analysis, we will consider
parameter points 
as providing a good description of
the excesses if they
account for the combined effect of the three excesses
at the level of $1\ \sigma$ or better.
For three independent measurements, this
translates into the requirement
\begin{equation}
\chi^2_{\gamma\gamma+\tau\tau+bb} 
\leq 3.53 \ .
\end{equation}

\section{Numerical analysis}
\label{sec:numana}

In this section we discuss our
numerical analysis, where we investigate whether 
the N2HDM type~IV parameter space that can describe 
the three excesses observed near $95\gev$
can also yield a predicted value for the $W$-boson mass,
$\MWNTHDM$, that is so 
large that it would be in agreement with the measured
value as recently reported by CDF.
We perform
a random scan in the N2HDM type~IV
over the free parameters
as defined in \refeq{eq:inputsnew}, where the scan
ranges were chosen to be
\begin{align}
&94 \gev \leq m_{h_1} \leq 98 \gev \; ,
\quad m_{h_2} = 125.09 \gev \; ,
\quad 300 \gev \leq m_{h_3} \leq 1000 \gev \; , \notag \\
&\quad 300 \gev \leq m_A \leq 1000 \gev \; ,
\quad 650 \gev \leq \MHp \leq 1000 \gev \; , \notag\\
&1 \leq \tb \leq 10 \; ,
\quad 0 \leq m_{12}^2 \leq 10^6 \gev^2 \; ,
\quad 100 \gev \leq v_S \leq 1500 \gev \; , \notag \\
&0.6 \leq c_{h_2 VV}^2 \leq 0.9 \; , \quad
0.6 \leq c_{h_2 t \bar t}^2 \leq 1.0 \; , \quad
\mathrm{sign}(R_{13}) = \pm 1 \; , \quad
-1 \leq R_{23} \leq 1 \; . \label{eq:ranges}
\end{align}
It should be noted here that we focus
our parameter scan on the parameter regions 
where $h_{95}$, corresponding to $h_1$, has a sufficiently large coupling to 
the gauge bosons and fermions 
so that it can give rise to the excesses that were observed in the Higgs
searches near $95\gev$.
Since $h_{95}$ obtains its couplings
to fermions and gauge bosons as a result of
the mixing with the
state $h_2$ that we identify with the observed Higgs boson 
$h_{125}$, we imposed an upper limit of
$c_{h_2 VV} \leq 0.9$ 
(see \citere{Biekotter:2022jyr}
for details).
We use the public code
\texttt{ScannerS}~\cite{Coimbra:2013qq,
Muhlleitner:2016mzt,Muhlleitner:2020wwk}, which
scans the parameters randomly over the given
range and applies the theoretical and experimental
constraints discussed in
\refse{sec:constraints}
(where we modified the
routines for the check against the EWPO).
We select parameter points with $\MWNTHDM$
within the $2\,\si$ confidence-interval
of the new CDF measurement given
in \refeq{cdf-new}.
We thus only take into account points which fulfill
\begin{align}
  \chi^2_{\MWCDFnew} = \frac{(\MWNTHDM - \MWCDFnew)^2}
      {(\De\MWCDFnew)^2} \le 4 \ ,
\label{eq:cdfcond}
\end{align}
with $\De\MWCDFnew = 0.0094 \gev$.

\begin{figure}[t] 
\centering
\includegraphics[width=0.6\textwidth]{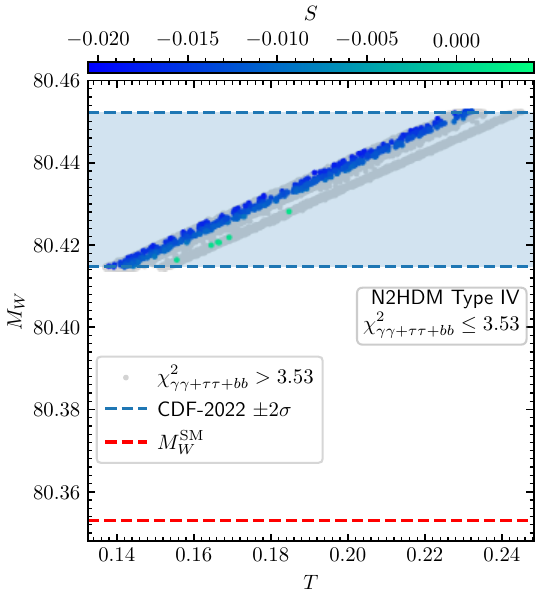}
\caption{\small
  The prediction for $\MW$ in the N2HDM as a function of $T$, where
  the color coding indicates the size of $S$ for the
  parameter points that describe the excesses at
  $95\gev$ at the level of $1 \ \sigma$ or better,
  i.e.~$\chi^2_{\gamma\gamma + \tau\tau + bb}
  \leq 3.53$. The remaining parameter points are shown
  in gray. The light blue region
  corresponds to the new CDF measurement within $\pm 2 \,\si$.
  All points lie within the light blue
  region, since in our scan we 
    selected the
  points fulfilling
  the requirement $\chi^2_{\MWCDFnew} \leq 4$.
  The red dashed line shows the SM prediction for $M_W$.}
\label{fig:T-MW}
\end{figure}

In \reffi{fig:T-MW} we show the
predicted values of $\MW$ in the N2HDM for the
parameter points of our scan 
as a function of
the oblique parameter~$T$.
For the parameter points that provide a description
of the collider excesses at $95\gev$ at the
level of $1\ \sigma$ or better,
i.e.~$\chi^2_{\gamgam+\tau\tau+bb} \leq 3.53$,
the color coding indicates the size of
the oblique parameter~$S$.
The remaining parameter points are shown in gray.
The light blue region
shows the new CDF measurement within its $\pm 2\,\si$
band. The
red dashed line indicates the SM prediction
for the $W$-boson mass (see
\refse{sec:int} for details).
According to \refeq{MW-STU},
one finds a nearly linear
dependence of $\MWNTHDM$ on $T$,
with a subleading contribution
coming from $S$.
The contributions from the
oblique parameter~$U$ are
found to be
negligible. The color coding indicates
that either values of
$S \approx -0.020$ (dark blue) or $S \approx 0$
but positive (green) are found.
The origin for the presence of these
two separate branches will be further discussed below.
It can be seen that the N2HDM of type~IV can
yield a prediction for $\MW$ that agrees with the new
$\MW$ measurement from CDF, while simultaneously providing a good
description of the three
excesses at about $95 \gev$, possessing a
Higgs boson at $125 \gev$ whose properties are compatible
with the LHC results, and which furthermore 
is in agreement with all the
other constraints listed in \refse{sec:constraints}. 

As a consequence
of the applied condition
shown in \refeq{eq:cdfcond}, we enforce in our 
scan that
all parameter points lie within the $2\ \sigma$
uncertainty band of the CDF measurement.
As such, our results demonstrate the compatibility
of a description of the excesses at $95\gev$ and
a prediction for $M_W$ that 
can be as large as
the CDF result. However, we stress
that parameter points that fit the collider
excesses do not necessarily predict a value for $M_W$
that would be inside the $2\ \sigma$ uncertainty band
of the CDF measurement.
In fact, parameter regions yielding a good 
description of the collider excesses near 95~GeV 
can also give rise to smaller predicted values 
of~$M_W$ that are closer to the SM value.
We would like to recall once more in this context 
that a potential new world average
value of $M_W$,
taking into account the CDF measurement and the
previous measurements at
LEP~\cite{ALEPH:2010aa},
the Tevatron~\cite{CDF:2013dpa}
and the LHC~\cite{ATLAS:2017rzl,LHCb:2021bjt},
is expected to have a central value between
the current world average and the CDF value.
One can extrapolate from the slope of the
line of points in \reffi{fig:T-MW} that
a parameter scan targeting such
a future
average value (once it becomes available)
would 
yield a preference for somewhat
smaller values of the
$T$ parameter. Besides that, our conclusions regarding
the compatibility of the description of the excesses
at $95\gev$ and a sizable positive shift
to $M_W$ in the direction of the CDF measurement
are not affected.

\begin{figure}[t] 
\centering
\includegraphics[width=0.6\textwidth]{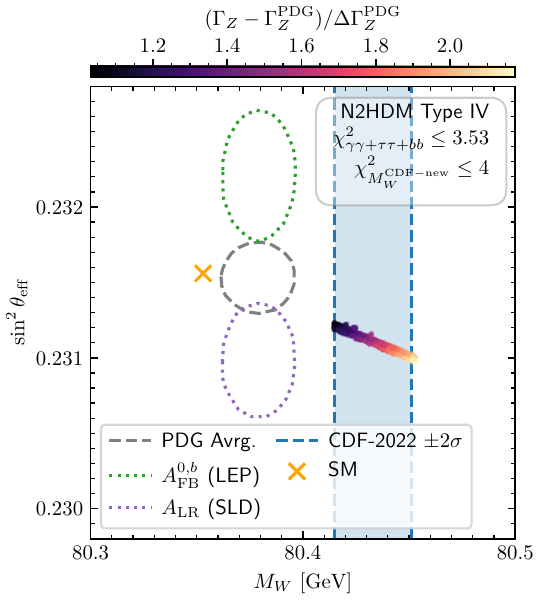}
\caption{\small
  The predictions for $\MW$ and $\sweff$ in the N2HDM.
  The color coding of the points indicates the
    difference between the prediction for
  $\Gamma_Z$ and the PDG average value
  $\Gamma_Z^{\rm PDG}$ divided by the
  experimental uncertainty
  $\Delta \Gamma_Z^{\rm PDG}$.
  The light blue region corresponds
  to the new CDF measurement within $\pm 2 \ \sigma$.
  The green dotted
  and the purple
  ellipses indicate
  the $68\%$ confidence level results
    from
  the two individually most precise 
  measurements of $\sin^2\theta_{\rm eff}$ via
  $A_{\rm FB}^{0,b}$ at LEP and
  $A_{\rm LR}$ at SLD, respectively,
  whereas the gray dashed ellipse
  indicates the 
    average of the LEP and SLD
  measurements~\cite{Heinemeyer:2004gx,
  Baak:2013fwa,ALEPH:2005ab}.
  Those ellipses are shown for the $\MW$ value corresponding to the current
  world average.
  The orange cross indicates the SM prediction.
}
\label{fig:MW-SW}
\end{figure}

Next we turn to the compatibility with 
the most sensitive $Z$-pole observable, $\sweff$.
In \reffi{fig:MW-SW} we show the predictions for $\MW$ and $\sweff$ in
the N2HDM for all points with
$\chi^2_{125} - \chi^2_{125,\mathrm{SM}} \leq 5.99$,
$\chi^2_{\gamgam+\tau\tau+bb}
\leq 3.53$ and 
$\chi^2_{\MWCDFnew} \le 4$.
One can see that the parameter points that fit the
new CDF measurement of the $W$-boson mass feature
also sizable modifications of $\sin^2\theta_{\rm eff}$
compared to the SM prediction.
It should be noted that without the restriction of the displayed points to
those for which the predicted value for $\MW$ is compatible at the 
$2 \ \sigma$ level with the new CDF measurement the
displayed scan poins in \reffi{fig:MW-SW} would extend further to the left,
i.e.\ into the direction towards the SM prediction that is indicated by an
orange cross in the figure.
According to \refeq{eq:sweffT}, the predicted
values of $\sin^2\theta_{\rm eff}$ featured
for the displayed parameter points of our scan are smaller
than the SM value, and they do not touch
the current $1\,\sigma$ ellipse
based on the PDG average values
of $M_W$ (which does not contain the new CDF measurement)
and $\sin^2\theta_{\rm eff}$.
However, here it should be kept in mind that the
PDG average of $\sin^2\theta_{\rm eff}$ is composed of 
different measurements, where the two most precise ones 
are compatible with each other only at
the level of about $3\,\sigma$: 
the one based on the forward-backward asymmetry 
of the bottom quarks
measured at LEP~\cite{ALEPH:2005ab}, and the one obtained from the 
left-right asymmetry
measured at SLD~\cite{ALEPH:2005ab}. It can be observed that the data points preferred by the $\MW$ measurement of CDF are in better
agreement with the SLD measurement based on
$A_{\rm LR}$, whereas the tension increases with
the value of $\sin^2\theta_{\rm eff}$ extracted
at LEP based on measurements of $A_{\rm FB}^{0,b}$.
Similar observations were made in \citeres{Bahl:2022xzi,
deBlas:2022hdk,Lu:2022bgw},
and the correlation between the effective weak
mixing angle and the mass of the $W$~boson
is expected to arise generically in models in
which the shift in the prediction for $\MW$ towards the
new CDF measurement of $M_W$ is
accommodated mainly via the breaking of the
custodial symmetry by means of a non-zero
$T$~parameter (and not via, e.g., BSM vertex and box
contributions to muon decay).
The fact that the points in \reffi{fig:MW-SW}
lie along an approximately straight line is
an indication of the strong dependence on
the $T$-parameter, whereas the impact of
the non-zero values of the $S$-parameter
on the prediction for
$\sin^2\theta_{\rm eff}$
is very small in our scan.

While the ellipse indicating the current world average for $\MW$ and
$\sweff$ as well as the two ellipses indicating the measurements of 
$\sin^2\theta_{\rm eff}$ via $A_{\rm FB}^{0,b}$ at LEP and
$A_{\rm LR}$ at SLD are all shown for the current world average of 
$\MW$ that does not contain the new CDF measurement, it becomes clear
from the plot that for
a future world average value of $\MW$ located in between the current
world average and the new CDF measurement the displayed ellipses would be
accordingly shifted to the right (and modified in order to account for the
combined experimental error of the new world average).
The tendency towards a better agreement of the N2HDM predictions with 
the SLD measurement of $\sweff$ based on $A_{\rm LR}$ will become more 
pronounced the closer the future world average for $\MW$
will be to the new CDF measurement.
Finally, it should be noted that,
while we only show in \reffi{fig:MW-SW} the
parameter points for which the excesses in the low-mass Higgs searches near
$95\gev$ are well described, the other parameter points 
that are in agreement with the CDF result for $\MW$
would be located at essentially
the same region in the $M_W$--$\sin^2\theta_{\rm eff}$
plane.\footnote{In
\citere{Biekotter:2021ovi} a complex
singlet field was considered
instead of the real singlet of the N2HDM.
There it was shown that the imaginary part of the
singlet can give rise to a valid dark-matter candidate
whose annihilation can also account for the
so-called galactic-center excess,
while the real component of the singlet
field gives rise to the state at $95\gev$
that accounts for the collider excesses as in the N2HDM.
Therefore, a combined description 
of the prediction for the $W$-boson mass, the excesses in the 
Higgs searches near $95\gev$ and
the galactic-center excess should
also be possible (see \citeres{Fan:2022dck,Zhu:2022tpr}
for discussions of the $\MW$ prediction and the galactic-center excess 
in the inert 2HDM).
}

Finally, we discuss the compatibility 
of the parameter region of the N2HDM yielding 
predictions for $M_W$ close to
the CDF measurement with
the experimentally measured
value of the width of the $Z$ boson, $\Gamma_Z$.
The color coding of the points in \reffi{fig:MW-SW}
indicates the difference between the prediction for
$\Gamma_Z$ and the PDG average value
$\Gamma_Z^{\rm PDG}$ divided by the
experimental uncertainty
$\Delta \Gamma_Z^{\rm PDG}$~\cite{ParticleDataGroup:2020ssz}.
Our N2HDM prediction for $\Gamma_Z = 
\Gamma_Z^{\rm SM} + \Delta \Gamma_Z$ is based
on the SM prediction computed
according to \citere{Freitas:2014hra} and using the input
parameters as discussed in \refse{sec:mw}.
The shift $\Delta \Gamma_Z$ is obtained
using the fit formula given in \citere{Burgess:1993vc},
where we took into account the numerically
relevant terms depending on $S$ and $T$.
One can see that a prediction for $M_W$ in
agreement with the $2\ \sigma$ uncertainty band
of the CDF measurement in combination with a
good description of the excesses at $95\gev$
is possible with a
tension for $\Gamma_Z$ which is only slightly
above $1 \ \sigma$.
For parameter points that predict a value of
$M_W$ that is even larger than the central
value of the CDF measurement, the tension for
$\Gamma_Z$ grows to the level of more than
$2 \ \sigma$.
Here we stress again that a future 
world average
value of $M_W$ taking into account the CDF
measurement would lie below the CDF measurement,
with potentially much larger uncertainties
reflecting the low level of compatibility 
between
the CDF result
and the 
most precise
other measurements of the $W$-boson mass.
Thus, an analysis based on such a future world
average value would
be expected to yield a prediction
for $\Gamma_Z$
that agrees with the experimental value
at the level of $1 \ \sigma$
or better.

\begin{figure}
\centering
\includegraphics[width=0.48\textwidth]{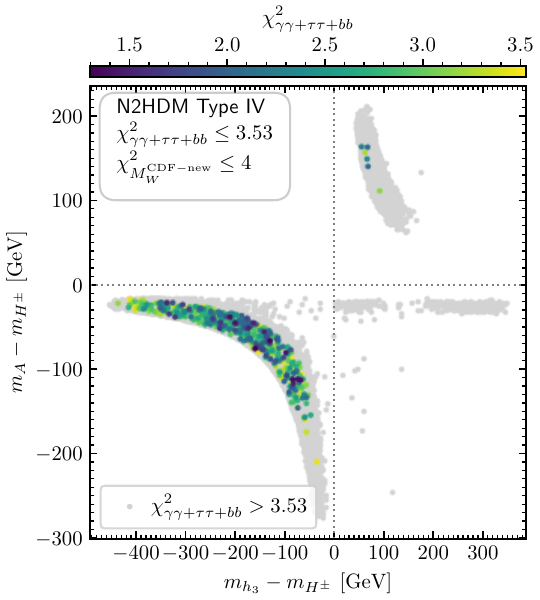}~
\includegraphics[width=0.48\textwidth]{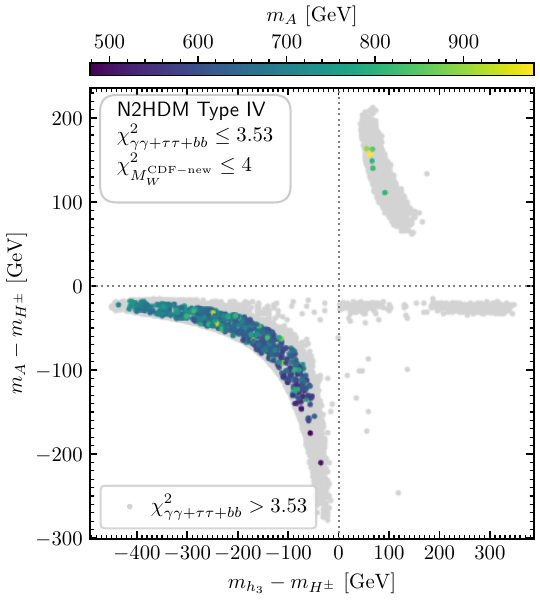}
\caption{\small
  Parameter points in the plane of the mass
  differences $m_{h_3} - m_{H^\pm}$ (horizontal
  axis) and $m_{A} - m_{H^\pm}$ (vertical axis).
  The color coding indicates the value of
  $\chi^2_{\gamma\gamma+\tau\tau+bb}$ in the
  left plot and of $m_A$ in
  the right plot for the parameter points
  that describe the excesses at
  $95\gev$ at the level of $1 \ \sigma$
  or better,
  i.e.~$\chi^2_{\gamma\gamma + \tau\tau + bb}
  \leq 3.53$. The remaining parameter points are shown
  in gray.
}
\label{fig:MW-masses}
\end{figure}

In order to shed
more light on the N2HDM parameter space regions
that can give rise to a value of $M_W$ that
is in agreement with the CDF measurement,
we show in \reffi{fig:MW-masses} the parameter points
in the plane of the mass differences between
$h_3$ and $H^\pm$ on the horizontal axis
and between $A$ and $H^\pm$ on the vertical
axis.
For the points with 
$\chi^2_{\gamgam+\tau\tau+bb} \le 3.53$,
the color coding 
indicates the value
of~$\chi^2_{\gamgam+\tau\tau+bb}$
in the left plot and of $m_A$
in the right plot.
The remaining points are shown in gray
(as throughout our analysis, the condition of \refeq{eq:chi125constraint} is applied for the displayed points).
One can see that 
most of the parameter points
have a mass hierarchy in which
$h_3$ and $A$ are either both lighter or
both heavier than the charged Higgs bosons $H^\pm$.
The separation of the parameter points into
these two distinct mass hierarchies that are
visible in \reffi{fig:MW-masses} is also the reason for the
presence of the two separate branches of
points that are visible in \reffi{fig:T-MW}.
The fact that the presence of these two mass hierarchies
facilitates a prediction of
the $W$-boson mass that is in agreement with the
recent CDF measurement is 
in line with the results of \citere{Bahl:2022xzi}, where the compatibility of the 2HDM of type I in the alignment limit with the CDF measurement of $\MW$ was investigated.
We note that in contrast to the results in the
alignment limit of the 2HDM, 
in the present analysis 
we also find 
parameter points with a predicted value of $\MW$ in agreement with the CDF measurement featuring a different mass hierarchy, namely
$m_A < m_{H^\pm} < m_{h_3}$,
as indicated by the displayed parameter points in the lower right parts of the plots in \reffi{fig:MW-masses}.
Most of these points --
which are all displayed in gray, indicating that they are not compatible with the excesses observed in the Higgs searches near $95\gev$
at the ${1 \ \sigma}$ level -- are located
on a horizontal branch
where $A$ and $H^\pm$ are close in mass.
For these points, the heavy state $h_3$ is
dominantly singlet-like.
Thus, this part of the N2HDM parameter space resembles the one of a 
2HDM with the doublet-like states $h_{95}$ and $h_{125}$ 
(scenarios of this kind, where the second-lightest CP-even Higgs boson corresponds to the observed state at about $125 \gev$, have been investigated in \citeres{Haisch:2017gql,
Bahl:2021str,Fontes:2017zfn})
which is supplemented by the heavier
singlet-like state $h_3$.
Consequently, in this case the large 
upward shift in the prediction of $M_W$ 
results from the large mass splitting
between $h_{95}$, $h_{125}$ and the states $A$ and $H^\pm$.
The fact that the points in the horizontal branch with 
$m_A \approx m_{H^\pm}$ are shown in gray
indicates that 
for a dominantly doublet-like state $h_{95}$ the three
excesses observed in the low-mass Higgs-boson searches 
cannot be described simultaneously.
We therefore do not discuss these points any further
in the following.

For the parameter points that accommodate the
excesses in the low-mass Higgs searches at the level of 
$\chi^2_{\gamma\gamma + \tau\tau + bb} \leq 3.53$,
one can see from the left plot of \reffi{fig:MW-masses} that both the
mass hierarchies
$m_{h_3},m_A < m_{H^\pm}$ and
$m_{h_3},m_A > m_{H^\pm}$ 
can be realized. For the points
in the upper right part of the plots for which
the charged Higgs bosons are lighter than
the heavy neutral states,
our scan resulted in only a
small number of parameter points
that comply with the various constraints.
While we expect that a more detailed scan would give rise to a 
somewhat larger allowed parameter region in the upper right part 
of the plots, we do not attempt a detailed discussion of the
correlation between $\chi^2_{\gamgam+\tau\tau+bb}$ and the
precise values of the mass splittings.
owever, it is obvious from the displayed gray points 
that even without imposing the constraint on 
$\chi^2_{\gamgam+\tau\tau+bb}$ the parameter region in this 
branch giving rise to a 
prediction for $\MW$ that is compatible with the new CDF value is more restricted than for the points
that feature $H^\pm$ as the heaviest particle 
(lower left branch of points).
Here it is important to note that
the mixing of the singlet field with the
doublet fields in the N2HDM
gives rise to additional
theoretical constraints on the scalar
couplings, in particular from perturbative
unitarity, as compared to the 2HDM.
As a result of these additional constraints
in combination with the requirement
$c_{h_{125}VV}^2 < 0.9$ (see \refeq{eq:ranges})
and experimental constraints from flavor-physics
observables,
we find that parameter points with
$m_{h_3} - m_{H^\pm} \gtrsim 150\gev$
in this branch
are excluded.

Turning now to the
points with the mass hierarchy 
$m_{h_3},m_A < m_{H^\pm}$ that are
visible in the lower left part of the plots,
one can see that 
the mass splitting between $H^\pm$ and $A$ is restricted to be below about 
$200 \gev$ for the points describing the excesses in the low-mass Higgs searches at the 
$1 \ \sigma$ level, while the mass splitting between $H^\pm$ and $h_3$
can be even larger than $400\gev$.
In the latter parameter region
$A$ and $H^\pm$ are almost
degenerate, while $h_3$ is substantially lighter.
The fact that this particular mass hierarchy
allows for a good description of the
excesses at $95\gev$ was already noted in
\citere{Biekotter:2022jyr} (see Fig.~5 therein).
It is remarkable
that the same mass hierarchy, as demonstrated here, can also give
rise to a prediction of $M_W$ in agreement
with the CDF measurement. 
Furthermore, for this mass hierarchy 
the strongest
first-order electroweak phase transition
can be accommodated 
in the (N)2HDM~\cite{Biekotter:2021ysx,Biekotter:2022kgf},
and as such the respective parameter space
regions might be suitable for the realization
of electroweak baryogenesis or
for the production
of an observable primordial
gravitational-wave background.

From the right plot of \reffi{fig:MW-masses} one can infer that 
in the lower left branch of points where $m_A$ is significantly 
smaller than $m_{h_3} \approx m_{H^\pm}$
the presence of a CP-odd Higgs boson
with a mass of $400\gev \lesssim m_A \lesssim 500\gev$
is compatible with a prediction of the $W$-boson
mass in agreement with the new CDF value and with
the presence of a Higgs boson at $95\gev$ that
is in fairly good agreement with the measured
signal rates in the three respective decay modes.
This is of particular interest in view of the fact 
that the local $3.5 \ \sigma$
excess observed at about $400\gev$ by
the CMS collaboration in searches for
additional Higgs bosons in di-top final
states~\cite{CMS:2019pzc}
can be described by means of the 
CP-odd Higgs boson 
$A$ of the type~IV N2HDM, while at the same
time the $\gamgam$ excess and the $b \bar b$
excess at $95\gev$ can be described by
the singlet-like Higgs boson 
as shown in \citere{Biekotter:2021qbc}.
The presence of parameter points with $m_A \lesssim 500\gev$,
as shown in the right plot of
\reffi{fig:MW-masses},
that also accommodate the more recently
observed $\tautau$ excess at $95\gev$
and the CDF measurement of $M_W$ leave room
for the possibility that additionally
also the $t \bar t$ excess at about $400\gev$
can be accommodated. We leave a more detailed
investigation of this 
very interesting scenario
for future work.

\section{Conclusions}
\label{sec:conclu}

The recently reported measurement of the
$W$-boson mass $M_W$ by the CDF collaboration
deviates significantly from the SM
prediction. 
A future world average for $\MW$ that includes the previous measurements of $\MW$ as well as the new CDF result will have to be based on a careful analysis of the systematic uncertainties of the different measurements and will have to account for the observed spread between the different results. If the future world average moves significantly towards the central value reported by the recent CDF measurement, the presence of sizable contributions from BSM physics to the prediction for $\MW$ would be favored. 
In this paper we have discussed whether a prediction for $\MW$ that would be compatible with the recent CDF measurement and simultaneously a description of recently reported excesses in low-mass Higgs boson searches near $95\gev$ can be accommodated in a simple extension of the SM without being in conflict with existing experimental and theoretical constraints. 
Specifically, we have focused on
a Two-Higgs-Doublet Model that is
extended with an additional real singlet
scalar (N2HDM) 
of Yukawa type~IV,
which we had identified in a recent
publication~\cite{Biekotter:2022jyr} to be suitable
for the description of the reported excesses in the search for 
light Higgs bosons at around $95\gev$.
These excesses were found in the $\gamgam$, the $\tau^+\tau^-$ and
the $b \bar b$ decay modes and can be described in the N2HDM 
by means of a singlet-like Higgs boson at this mass
that mixes with the detected Higgs boson
at $125\gev$.

We have demonstrated that in the
N2HDM a prediction of the value
of $M_W$ in agreement with the CDF measurement
is compatible with a good description
of the collider excesses at $95\gev$.
We have shown that the parameter regions giving rise to those features 
are in agreement with
the various theoretical and experimental constraints
on the model parameters.
In the parameter regions that we have determined
the heavy neutral scalars
$h_3$ and $A$ are either both lighter or both
heavier than the charged states $H^\pm$.
Furthermore, we have pointed out that the parameter region 
featuring a sizable
mass splitting between the third CP-even
Higgs boson $h_3$ and the approximately
mass-degenerate CP-odd Higgs boson $A$ and the
charged Higgs bosons $H^\pm$,
i.e.~$m_{h_3} < m_A \approx m_{H^\pm}$
(with $m_A < m_{H^\pm}$), naturally
gives rise to sizable values
of the $T$ parameter and thus to an enhancement
of $M_W$ compared to the SM prediction.
It is interesting to note that this is also
the mass hierarchy which is favoured by
the requirement of realizing a
strong first-order
electroweak phase transition 
(which may give rise to an observable gravitational wave signal)
in the N2HDM.
Concerning the effective weak mixing
angle, the parameter space preferred by the new $\MW$ measurement 
from CDF is 
well compatible with the $\sweff$ value extracted from 
$A_{\rm LR}$ at SLD, 
while there is some tension
with the value obtained from $A_{\rm FB}^{0,b}$ at LEP.

Finally, we note that a possible future 
world average
value, taking into account the previous measurements
of $M_W$ at LEP, the Tevatron and the LHC in
combination
with the new CDF result, would 
be expected to have a somewhat lower 
central value for $M_W$  than the one reported by 
CDF
and potentially a significantly larger
uncertainty 
reflecting the low level of compatibility
between
the CDF result and the most precise previous 
measurements.
Such a central 
value of
$M_W$ 
between the current world average and the CDF value
would 
yield a preference for
smaller and 
somewhat less 
restricted
values of the $T$~parameter. These smaller values of
$T$ would correspond to slightly smaller
mass splitting of the heavy BSM Higgs bosons.
We stress that our qualitative results indicating
the compatibility
of the description of the collider excesses at
$95\gev$ with a sizable upward shift 
in the
prediction for $M_W$ would remain unchanged 
in such a scenario.

\subsection*{Acknowledgements}

The work of S.H.~is supported in part by
the grant PID2019-110058GB-C21 funded by
``ERDF A way of making Europe'' and by
MCIN/AEI/10.13039/501100011033, and in part
by the grant CEX2020-001007-S funded by
MCIN/AEI/10.13039/501100011033.
The work of T.B.~and G.W.~is supported by the Deutsche
Forschungsgemeinschaft under Germany’s Excellence Strategy EXC2121
“Quantum Universe” - 390833306.
This work has been partially funded by the Deutsche Forschungsgemeinschaft 
(DFG, German Research Foundation) - 491245950.

\bibliography{lit}

\end{document}